\documentclass[conference]{IEEEtran}
\usepackage{nopageno}
\makeatletter
\def\footnoterule{\relax%
  \kern-5pt
  \hbox to \columnwidth{\hfill\vrule width 0.8\columnwidth height 0.4pt\hfill}
  \kern4.6pt}
\makeatother
\ifCLASSOPTIONcompsoc
\usepackage[nocompress]{cite}
\else
\usepackage{cite}
\fi
\usepackage{wrapfig}
\usepackage{microtype}
\usepackage{color, soul}
\usepackage{booktabs}
\usepackage{enumitem}
\usepackage{algpseudocode}

\usepackage[table]{xcolor}

\usepackage[table]{xcolor}
\usepackage{tabularx}
\usepackage{amsmath}
\usepackage{algorithm,algorithmicx,algpseudocode}

\usepackage{syntonly}
\usepackage{multicol}
\usepackage{array}
\usepackage[tight,footnotesize]{subfigure}
\usepackage{stfloats}
\usepackage{url}
\usepackage{gensymb}
\usepackage{multirow}
\usepackage{amsmath,array,graphicx}
\usepackage{tabularx}
\usepackage{footnote}
\usepackage{courier}
\usepackage{amssymb}
\usepackage{pifont}
\usepackage{tikz}
\newcommand{\xmark}{\ding{55}}%
\usepackage{todonotes}
\usepackage{etoolbox}
\usepackage{etoolbox}
\usepackage{tikz}
\usetikzlibrary{tikzmark}
\usetikzlibrary{calc}

\errorcontextlines\maxdimen

\begin{document}
\title{\Huge{Software Variants for Hardware Trojan Detection and Resilience in COTS Processors}}

\author{
    \IEEEauthorblockN{Mahmudul Hasan\IEEEauthorrefmark{1}, 
    Jonathan Cruz\IEEEauthorrefmark{2}, Prabuddha 
   Chakraborty\IEEEauthorrefmark{2}, 
     Swarup~Bhunia\IEEEauthorrefmark{2}
    and Tamzidul~Hoque\IEEEauthorrefmark{1}}
    \IEEEauthorblockA{
    \IEEEauthorrefmark{1}Department of Electrical Engineering and Computer Science,   University of Kansas 
    \\ \{m.hasan, hoque\}@ku.edu}
    \IEEEauthorblockA{\IEEEauthorrefmark{2}Department of Electrical and Computer Engineering,   University of Florida  \\\{jonc205,   p.chakraborty\}@ufl.edu, swarup@ece.ufl.edu }
    
   }

\maketitle

%
\IEEEpeerreviewmaketitle

\begin{abstract}
	The commercial off-the-shelf (COTS) component based ecosystem provides an attractive system design paradigm due to the drastic reduction in development time and cost compared to custom solutions. However, it brings in a growing concern of trustworthiness arising from the possibility of embedded malicious logic, or hardware Trojans in COTS components. Existing trust-verification approaches, are typically not applicable to COTS hardware due to the absence of golden models and the lack of observability of internal signals. In this work, we propose a novel approach for runtime Trojan detection and resilience in untrusted COTS processors through judicious modifications in software. The proposed approach does not rely on any hardware redundancy or architectural modification and hence seamlessly integrates with the COTS-based system design process. Trojan resilience is achieved through the execution of multiple functionally equivalent software variants. We have developed and implemented a solution for compiler-based automatic generation of program variants, metric-guided selection of variants, and their integration in a single executable. To evaluate the proposed approach, we first analyzed the effectiveness of program variants in avoiding the activation of a random pool of Trojans. By implementing several Trojans in an OpenRISC 1000 processor, we analyzed the detectability and resilience during Trojan activation in both single and multiple variants. We also present delay and code size overhead for the automatically generated variants for several programs and discuss future research directions to reduce the overhead. 
\end{abstract}

\begin{IEEEkeywords}
	Hardware Trojans; COTS; Trojan Resilience
\end{IEEEkeywords}
\section{Introduction}
\label{sec:intro}
\thispagestyle{empty}
In recent years, commercial off-the-shelf (COTS) electronic components have seen increased adoption in diverse domains, including military, avionics, finance, and commercial applications. Since designing a custom hardware product often results in longer deployment period with significant design and manufacturing costs, the use of COTS components prevails even in the most sensitive electromechanical systems such as NASA spacecrafts~\cite{ref:nasa}. 
However, adopting the COTS flow does not come without its own problems.
Security and reliability challenges arise from the fact that COTS component development involves external design and manufacturing facilities. Any of these supply-chain entities could introduce a hidden malicious modification or hardware Trojan in the design to cause a  functional failure or leakage of secret information (e.g., encryption keys) during field operation. 
Existing research on hardware Trojans largely focuses on addressing Trojan insertion in two major supply-chain entities: untrusted foundry and untrusted hardware intellectual property (IP) vendors. 
Techniques to detect Trojans inserted in an Integrated Circuit (IC) by comparing it with trusted or reference design have been investigated extensively for almost a decade \cite{xiao2016hardware}. Unfortunately, these post-silicon detection techniques do not apply to COTS components due to the following reasons: 
\begin{itemize}
\item For in-house ICs, a golden design is available to the design house, which does not apply to COTS IC.
\item Malicious COTS component developers have more flexibility in implementing complex Trojans compared to adversaries at the foundry, since, area, power, and delay constraints cannot be strictly defined during procurement.   
\end{itemize}

Additionally, researchers have proposed several static and dynamic analysis based Trojan detection methods for untrusted soft-IPs obtained from  third-party IP (3PIP) vendors. 
However, they cannot be applied for COTS ICs due to following:
\begin{itemize}

\item  Approaches for 3PIPs that apply static analysis on the netlist, such as boolean functional analysis \cite{ref:fanci} and machine learning~\cite{hoque2018hardware2} require white-box accessibility to design. Application of these techniques to COTS ICs would require destructive reverse engineering of the netlist.  
\item Dynamic approaches that detect Trojans through logic simulation are also not applicable, since the gate-level netlist is often required for the generation of test patterns.   
\end{itemize}

Hence, authentication of COTS electronics is certainly one of the most difficult trust-verification challenges and has not received the attention it needs \cite{xiao2016hardware}. 
A very few ideas have been developed in enabling Trojan resilient computing, specifically in general purpose COTS processors. In the SAFER PATH technique \cite{ref:safer}, execution of the same program in multiple processing units acquired as individual COTS component from different vendors collectively ensures Trojan resilient computation. This framework requires integration of multiple processing units and various hardware level modifications.

In this paper, we propose a novel and purely software-based solution to enable Trojan detection and resilience in COTS processors during field operation against Trojans that corrupt the program outputs (i.e., modification of variables). 
During the software integration step of the COTS processor, the program to be executed is divided into several segments of codes (defined as compare-blocks). These blocks define the boundaries at which the program states are to be checked during field operation. The code of each compare-block is transformed to a sequential execution of two ``variant codes'' that are each functionally equivalent to the original code. The program state (relevant variables) after the completion of the first variant is stored temporarily and compared with the resultant state after the second variant is executed. The key idea is that significant differences in frequency and sequence of both opcode and operands among the functionally equivalent variants results in \textit{dissimilar switching activities of the internal nets during the execution of individual variants}. Therefore, even if a hard to activate Trojan gets triggered in one of the variants and impacts the program states, it will not re-trigger in the same way in any other variant. Even if it does, the impact of the payload (i.e., corruption of critical variable) may not be identical. Dissimilar program states resulting from subsequent executions of two functionally equivalent codes indicate either a Trojan activation or fault propagation. Only during this rare instance, a third variant is executed that is significantly different than the earlier variant codes. A majority voting is executed among the three executions (i.e. variant 1, 2, and 3) to ensure the integrity of the computation even under possible activation of Trojan(s) causing a transient corruption. 
The proposed solution is distinctive from the other Trojan resilience techniques \cite{ref:safer} due to the following:
\begin{itemize}

 \item It does not depend on redundant processing units or architectural modifications to accommodate voting or synchronization. This translates to shorter time-to-market, lower development cost, and logistical flexibility.      

\item This can be applied to legacy COTS processors already deployed in the field. 

\item Since it does not depend on multiple cores or processing elements, acquisition of cores from diverse vendors to evade similar Trojans in all cores is not required. 
\end{itemize}

Please note that we can even run more than three variants (e.g. 5, 7, 9) to improve the probability of withstanding the Trojan without impacting the regular performance, as only two variants would be executed at runtime and others would be invoked upon Trojan activation. 
Generating a large number of variants of a given program is a non-trivial task. To automatically  generate the variants, we have proposed a methodology and tool flow for using existing compiler infrastructure that will require minimal tool development effort and can  generate variants for most processor architectures available in today's market. 
Overall, we make the following major contributions in this paper:

\begin{itemize}
	\item We propose a software-based framework that enables Trojan detection and resilience in COTS processors without any design or assembly-time hardware modification. Our technique is one of the very few that are directly applicable to COTS processors. 
	\item We present an automatic variant generation tool flow using LLVM compiler infrastructure that generates 
	the diverse variants using a code similarity metric. 
	\item By analyzing the signal transition data inside a 32-bit RISC microprocessor, we have observed the effectiveness of the tool-generated variants in avoiding activation of the same Trojan in multiple variants. 
	
	\item We have observed the detection and resilience of the variants against various combinations and sequential Trojan implemented inside the OpenRISC \textit{mor1kx} CPU \cite{mor1kx}. 
	We were able to detect and tolerate
	\footnote{In our context, tolerance means the ability to provide the correct output under Trojan activation. However, we define our protection as \textit{resilience} \cite{tolerance}, since a Trojan attack only on specific data in the whole system is tolerated.} 
	majority of the Trojans causing transient corruption.  
	
	\item We present delay and code size overhead of our approach by generating the variant-integrated code for several C programs in the MiBench suite. 
	
\end{itemize}

The rest of the paper is organized as follows. Section \ref{sec:pril} briefly describes our threat model. Section \ref{sec:method} describes our methodology and tool flow. 
We elaborates the effectiveness of our approach against Trojans and the corresponding overhead in Section \ref{sec:result}. Furthermore, we provide discussion on possible attacks and multi core execution of variants in Section \ref{sec:disc}. Finally, we conclude in Section \ref{sec:conc}. 

\section{Background and Preliminaries}
\label{sec:pril}

\subsection{Threat Model}
\textbf{Trojan Insertion Phase:}
From the third-party IP vendor to the foundry, we assume that a malicious modification of the design can be made at any step of the COTS processor design and manufacturing flow.

\textbf{Trojan Location:}
Even though our framework aims to bypass a Trojan within the processor core, a Trojan residing in the cache memory may not be covered. Due to the lack of flexibility in controlling physical read-write location within the cache, such Trojans can experience the desired triggering inputs among multiple variants. 

\textbf{Payload:}
Our framework focuses on Trojans that corrupts the state of certain registers or wires within the processor that leads to a targeted change in the output (i.e., certain variables) of a program. Such targeted change in program output can be performed by a Trojan that causes a transient corruption once triggered, instead of a persistent corruption. 
We are not considering a Denial-of-Service (DoS) attacks that simply prevents the normal execution of the program. 


 \begin{table}[!t]
 \begin{center}
        \scriptsize\addtolength{\tabcolsep}{-2.5pt}
\renewcommand{\arraystretch}{1}
\centering
\caption{Applicability of the Proposed Approach Together with Logic Testing in Detecting and Tolerating a Broad Class of Trojans}
\label{trojan_class}

\begin{tabular}{c|c|c|c|c} 
\toprule
\hline
Class   &  Coverage in & Coverage in   & Primary Trojan  & Example \\
 		 &  Logic Testing    & Proposed            & Property    &                \\ \hline
 1       & High                 & Low                 & Simple trigger     & Single opcode triggered           \\ 
         &                    &                     &  condition &          combinational HT       \\ \hline
 2      & Low                & High               & Extremely rare  &   Triggered by long  \\ 
         &                    &                &     trigger condition   &  sequence of opcode\\ \hline
 3      & Uncovered           & Uncovered & Trig. independent, & HT draining battery,      \\ 
         &                   &      &   incomparable payload  &    always-on leakage HT                 \\ \hline

\bottomrule
 \multicolumn{3}{l}{HT= Hardware Trojan}

\end{tabular}
 \end{center}
 \end{table}

\textbf{Trigger Mechanism:}
We assume an attacker designs a Trojan based upon some rare activation condition within the COTS IC. 
The trigger mechanism can be combinational or sequential. 
If a Trojan triggers in multiple variants and corrupts the program states of each variant similarly, it would not get detected. However, if the activation condition is rare, it is unlikely that the same Trojan will trigger in multiple variants.
We broadly classify the large functional trigger space in Table \ref{trojan_class}. Class-1 contains Trojans with simple triggering conditions that are hard to address using our method but can be detected during traditional IP verification phases. Hence, we do not necessarily need to cover them if a directed pre-deployment verification step is present. However, complex trigger mechanisms are likely to activate during logic testing (Class 2) to avoid detection. Such difficult-to-trigger Trojans are better addressed using our proposed method. Trojan activation based on parametric features (voltage, temperature), memory access, page faults, etc., are not considered and are beyond the scope of this work.

\subsection{Existing Work}


\subsubsection{Countermeasures for In-house ICs and 3PIPs}
The initial era of hardware Trojan research was focused on post-silicon and run-time detection of foundry inserted Trojans in ICs. As shown in Table \ref{existing}, these techniques are mostly either side-channel analysis based \cite{hoque2017golden, rad2008sensitivity, jin2008hardware, soll2014based, forte2013temperature} or logic-testing based approaches \cite{chakraborty2009mero, banga2008region}. Almost all of these techniques rely on one or more of the following requirements: 1) a golden design, 2) golden side-channel signature, and 3) design-time modifications (to implant sensors or to increase observable/controllable points). None of the above-mentioned requirements can be met in COTS IC based development model, rendering such techniques inapplicable. The requirement of design-time modification also eliminates design-for-security \cite{cao2014cluster, jin2012post, xiao2015efficient}, hardware obfuscation \cite{chakraborty2009security, dupuis2014novel, yasin2015transforming}, and split manufacturing \cite{vaidyanathan2014efficient, imeson2013securing} based solutions that are considered effective in preventing Trojan insertion at the untrusted foundry.

\begin{table*}[t!]
\centering
\caption{Comparison with Existing Countermeasures against Hardware Trojans}
\label{existing}
\resizebox{.9\textwidth}{!}{%
\begin{tabular}{c|c|c|c|c|c}
\hline
Approaches                             & Reference                                                                                   & Detection~/      & HT         & HT         & HT in      \\
                                       &                                                                                        & Tolerance Phase  & in IP      & in IC      & COTS IC    \\ \hline
\cellcolor{gray!75} Software Variant   & Proposed Approach                                                                    & Runtime          & \checkmark & \checkmark & \checkmark \\ \cline{2-6} 
\cellcolor{gray!75} IC Redundancy      & SAFER PATH \cite{ref:safer}                                                            & Runtime          & \xmark     & \checkmark & \checkmark \\ \cline{2-6} 
\cellcolor{gray!75} IP Redundancy      & TrojanGuard \cite{malekpour2017trojanguard}, HLS \cite{rajendran2016building}          & Runtime          & \checkmark & \xmark     & \xmark     \\
\cellcolor{gray!75}                    & Task Scheduling \cite{ref:mpsoc}, \cite{wang2018security}                              &                  &            &            &            \\ \hline
\cellcolor{gray!25} Static Analysis    & HAL \cite{fyrbiak2018hal}, COTD \cite{ref:COTD}                                           & Pre-Silicon      & \checkmark & \xmark     & \xmark     \\  \cline{2-6} 
\cellcolor{gray!25} Formal Methods     & Proof Check \cite{love2012proof}, Verification \cite{rajendran2015detecting}           & Pre-Silicon      & \checkmark & \xmark     & \xmark     \\ \cline{2-6} 
\cellcolor{gray!25} IP Monitoring      & Many-core \cite{kulkarni2016svm}, ISA Power \cite{lodhi2017power}                      & Runtime          & \checkmark & \xmark     & \xmark     \\ \cline{2-6} 
\cellcolor{gray!25} Side-Channel       & Power \cite{hoque2017golden}, \cite{rad2008sensitivity}, Delay \cite{jin2008hardware}, & Post-Silicon,    & \xmark     & \checkmark & \xmark     \\ \cline{2-6} 
\cellcolor{gray!25} Logic Testing      & MERO \cite{chakraborty2009mero}, Region Based \cite{banga2008region}                   & Pre/Post-Silicon & \xmark     & \checkmark & \xmark     \\ \hline
\cellcolor{gray!5} Design-for-Sec.     & Sensors \cite{cao2014cluster}, \cite{jin2012post}, OBISA \cite{xiao2015efficient}      & Pre/Post-Silicon & \xmark     & \checkmark & \xmark     \\ \cline{2-6} 
\cellcolor{gray!5} Hardware  obfuscation          & Logic Encryption \cite{dupuis2014novel},                                               & Pre-Silicon      & \xmark     & \checkmark & \xmark     \\ 
\cellcolor{gray!5}     &  HARPOON \cite{chakraborty2009security}, Camouflaging \cite{yasin2015transforming}       &   &   &   &     \\  \cline{2-6}
\cellcolor{gray!5} Split Manufacturing & \cite{vaidyanathan2014efficient}, 3D integration \cite{imeson2013securing}             & Pre-Silicon      & \xmark     & \checkmark & \xmark    \\ \hline
\multicolumn{5}{l}{Class of Countermeasure:
                 ~~\tikz\draw[black,fill=gray!85] (0,0) circle (.8ex); Trojan Tolerance,
                 ~~\tikz\draw[black,fill= gray!45] (0,0) circle (.8ex); Trojan Detection,
                 ~~\tikz\draw[black,fill= gray!10] (0,0) circle (.8ex); Trojan Prevention

}\\ 
\end{tabular}%
}
\end{table*}

More recently, several novel approaches have been developed to detect or tolerate untrusted design house inserted Trojans in soft-IPs (RTL/gate-level netlist). Detection oriented solutions like static analysis of the code \cite{ref:fanci, ref:COTD, banga2010trusted, zhang2015veritrust} and application of formal methods \cite{love2012proof, rajendran2015detecting} require a white-box accessibility of the hardware IP. Trojan tolerance  through IP redundancy \cite{malekpour2017trojanguard,rajendran2016building} and detection through run-time monitoring \cite{kulkarni2016svm, lodhi2017power} further demand modification of the hardware code, and/or IP vendor diversity. Since there is no flexibility to introduce design-time modification or analysis of the IP during COTS processor development, such solutions are also inapplicable.    

\section{Methodology}
\label{sec:method}
\begin{figure}[t!]
\centerline{\includegraphics[width=\columnwidth]{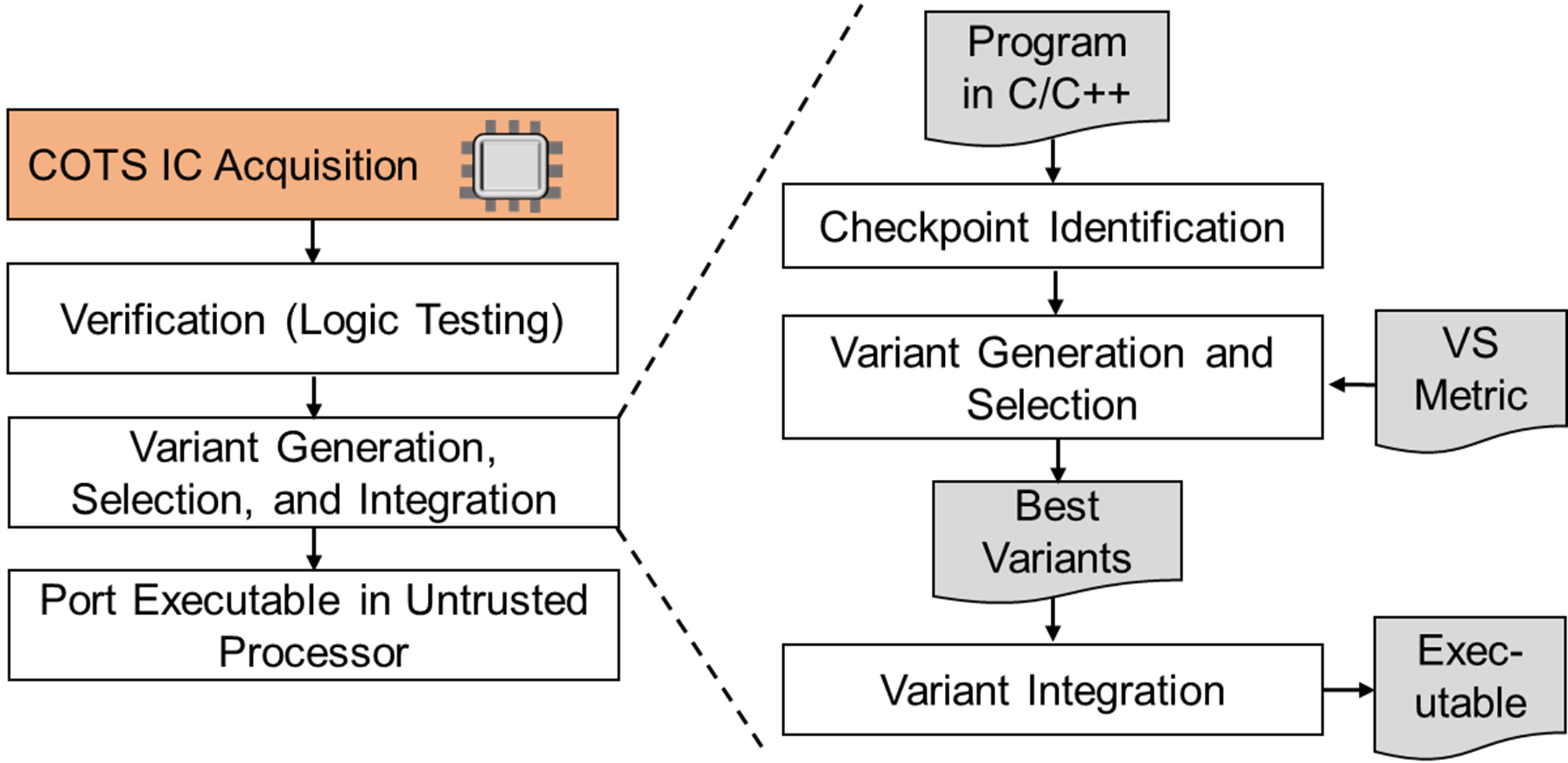}}
\caption{Overall flow of the proposed framework: A processor acquired from COTS supply chain is initially tested using logic testing to detect easy to trigger Trojans. Next, the software program to be executed on the processor is transformed to several variants and the  most diverse variants are selected using $VS$ metric and integrated in a single executable. }
\label{fig:toolFlow}
\end{figure}

An overview of the proposed approach is presented in Fig. \ref{fig:toolFlow}. For each checkpoint of the program where we want to verify the correctness of the execution, our goal is to generate a set of diverse variants based on a metric  called \textit{Variant Similarity (VS)}. The code of the generated variants are different from each other in code size, instruction and operand order and frequency, but functionally equivalent. 
Once the diverse variants are generated, they are integrated such that two variants are executed in a sequence and their results are compared. ``Result'' could be the outputs of certain computations in the  program as defined by the user (e.g., variables that are considered critical). The comparison of variant's outputs facilitates runtime Trojan detection. To facilitate Trojan resilience, three or more variants must be executed for majority voting. Therefore, along with the two variants, we integrate additional codes for conditional execution of three or more variants. The code of integrated execution of all variants will replace the original segment of the program from which variants are generated. This process is repeated for all checkpoints in the program as defined by the user. Finally, a single executable is generated from the transformed code. 
Below we describe these steps in detail.

\subsection{Verification}
The COTS processor deployment process is suggested to have a thorough verification step due to the potential trust and reliability issues \cite{koch2012role}. 
This verification step is important since it complements our proposed solution. As discussed using Table \ref{trojan_class}, the proposed solution has a higher probability of thwarting Trojans with very rare trigger conditions that could be hard to activate using logic testing within a limited time budget. Hence, this step should apply maximum possible test vectors within the limited test time or other verification techniques to detect any easy to activate Trojans.

\subsection{Checkpoint Identification}
Before applying any form of transformation to the code, it is necessary to realize the boundaries of variant execution defined as compare-blocks within a program. The end of each block is a checkpoint for comparing the results of a variant pair. The boundary can be within the range of any arbitrary block size where computation can be paused for comparison before moving to the next process. If the user is concerned about Trojan activation in a specific critical region of the code, the compare-block could be described for that region alone. In our current implementation, we support individual functions as compare-blocks.

\subsection{Metric}

\begin{figure}[t!]
\centerline{\includegraphics[width=.9\columnwidth]{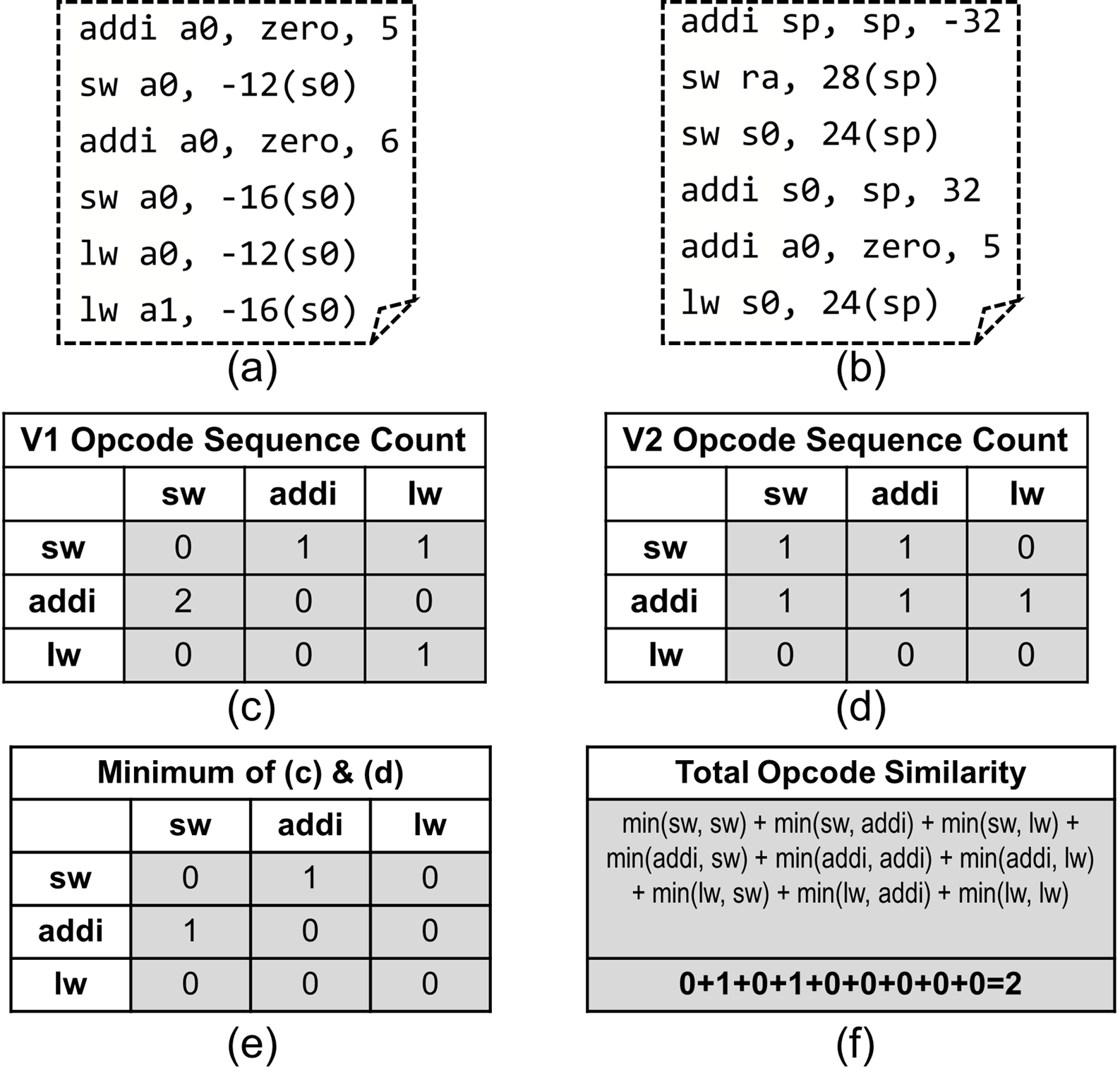}}
\caption{Example of variant similarity calculation for opcode segment among two variants V1 and V2. Figure (a) and (b) shows assembly codes of  variant V1 and V2 respectively, (c) and (d) shows  opcode sequence count table for V1 and V2, (e) shows the minimum of the counts from the two tables, and (f) shows the total opcode similarity.  }
\label{fig:vs}
\end{figure}

To estimate the susceptibility  of a pair of variants in activating the same Trojan, the $VS$ metric is formulated. Any two variants of a program ideally should have a diverse frequency and sequence of opcode and operands such that when they execute, the internal activity within the process is very different. Different switching activity of the internal nets during the execution of two variants will significantly decrease the probability of a Trojan being activated in both variants at the same point of execution. If a Trojan is activated only in one variant, the likelihood of detecting the Trojan increases. Even if the same Trojan gets activated in both variants, the impact of the Trojan payload will most likely be different due to the variant code diversity. We prefer to calculate similarity as the case of maximum similarity among two programs are known (i.e., when the programs are identical), which serves as a reference in understanding the value of VS. Besides, measurement of similarity has been well-researched in other domains \cite{runwal2012opcode}.  Among the large number of variants, the pair that provides minimum similarity can be considered to be most diverse. Quantification of code similarity has received great attention due to its use in metamorphic malware detection \cite{zhang2007metaaware, runwal2012opcode}. We leverage the technique presented in \cite{runwal2012opcode} to calculate code similarity among two variants. To calculate $VS$, we first segment each instruction of a defined code segment into opcode and operand section. Next, we calculate the counts of all consecutive opcode pairs for each variant as shown in Fig. \ref{fig:vs}. The dimensions of the table are $N \times N$, where $N$ is the number of opcodes available in the instruction set architecture. While \cite{runwal2012opcode}, only performs opcode analysis, we also generate similar count tables for instruction operands as we want to generate variants that are diverse with respect to both instruction and data. The opcode and operand count tables for each variant are flattened and appended as a single vector. We compute $VS$ by aggregating the element-wise minimum between vectorized opcode and operand  sequence count tables for variant pairs as shown in Fig. \ref{fig:vs}.


Please note that under multithreading and out of order execution, the order of instructions executed at runtime could differ from the sequence present in the assembly code being analyzed for calculating $VS$. However, the similarity is observed over all consecutive instruction pairs (instead of long sequence of instructions) that are less impacted by such change in execution order. 
Moreover, if register renaming is supported, the architectural registers that are present at the assembly program are not the same as the physical registers. This could make the operand part of $VS$ calculation unreliable. In such a scenario, we can disregard operand similarity for $VS$ calculation and only utilize the opcode segment.


\subsection{Compiler-based Automatic Variant Generation}
We leverage existing compiler infrastructure that will require minimal tool development effort and can generate variants for most processor architectures available in today's market.  Modern compilers provide various optimization passes that transform the code to obtain faster runtime and code size. A large number of code transformations are possible using different optimization passes. Therefore, by applying various combinations and sequences of optimization passes, we can generate a large number of variants. From this large pool of variants, we can identify the best variants using the $VS$ metric.

\begin{figure}[t!]
\centerline{\includegraphics[width=\columnwidth]{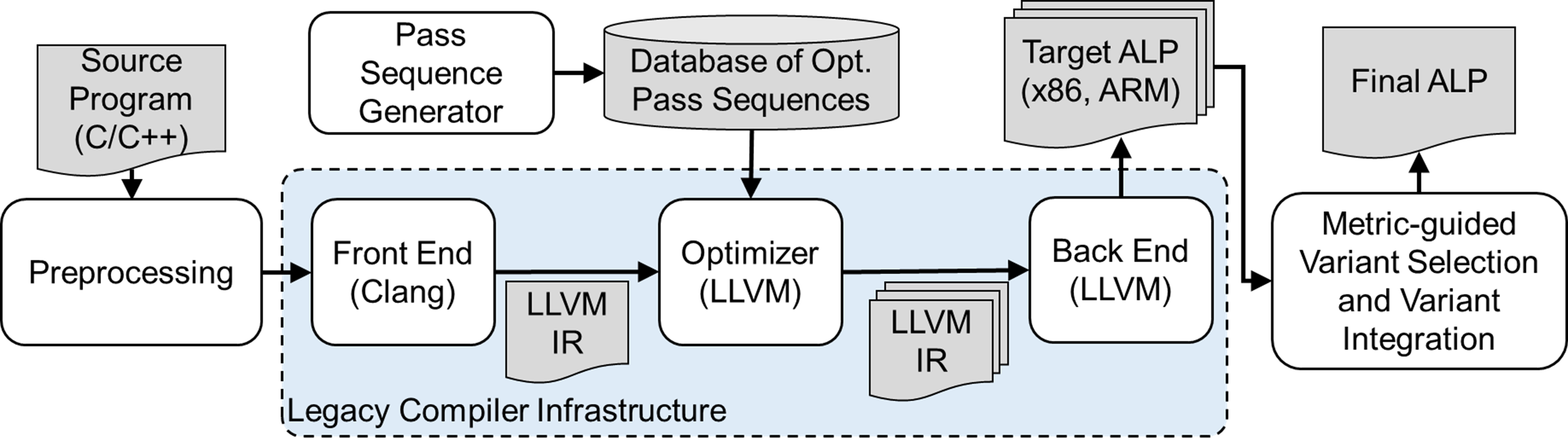}}
\caption{Automatic variant generation flow using legacy compiler infrastructure. }
\label{fig:llvm}
\end{figure}

Fig. \ref{fig:llvm} shows the toolflow for the compiler infrastructure using LLVM and Clang in a dotted box. This infrastructure uses a three phase modular design process (i.e., front end, optimizer, and back end) that is amenable to modification. The front end takes a high level source code like C/C++ and maps it to an architecture independent LLVM Intermediate Representation (IR). The LLVM optimizer can apply various transformations to the IR using a series of optimization passes. A user can select a custom series of passes to apply certain transformations and generate a modified IR that is likely to be optimized with respect to code size or runtime. Therefore, by applying different sets of passes to the same IR, we can potentially generate diverse IR codes that are functionally equivalent. When transformed to assembly programs (ALPs), these IR codes are likely to have different usage of opcodes an operands. Using the $VS$ metric, we can assess all the ALPs and select the variants with lowest similarity among each other.

 LLVM contains a large number of distinct passes that leads to infinite number of possible optimization sequences \cite{jain2014finding}. Therefore, application of all possible pass sequences to find the most diverse variants is not feasible. Prior work on compiler pass selection has focused on the generation of pass sequences for various optimization goals (e.g., code size, power, execution time) using heuristic based search techniques and machine learning. Hence, identification of  a database of minimal pass sequences for generating the most diverse set of codes would be a future research. In this work, we used a database of pass sequence generated from the ones presented in \cite{ashouri2017micomp, jain2014finding}.



\subsection{Integration of Variants}
\label{integration}
\texttt{system()} is a standard function  in C that executes a specified command by calling the command processor (e.g., UNIX shell or CMD in Windows). The \texttt{system()} function can direct a given program (say host program) to run another program (say external program). After the external program is executed, the control comes back to host and the statements following the \texttt{system()}function continues to execute. We can use the \texttt{system()} function to seamlessly integrate the generated variants. To store the results of each variant we can use \texttt{fprintf()} function in C. Let us assume that there is a compare-block within a given program and we have generated executable of three variants for that block (e.g., v1, v2, v3). To integrate the variants, first we would remove the original compare-block code. In place of the compare-block, we will call the first variant using \texttt{system(v1)}, followed by \texttt{fprintf(filename, variable)} to store the state of relevant variable in a file. We will similarly run the second variant and after that additional codes will be included to read-back the two files and compare the variable states among the variants. Following this method, we can run any number of variants and compare their results. 

However, because these functions are integrated \textit{after} the variant generation process, their assembly codes could be identical when used with different variants. An attacker could potentially bypass detection by designing a Trojan that activates based on the execution of these known functions, instead of the original program, allowing the attacker to design a Trojan that activates after each variant (when the identical system call functions are used). To prevent this vulnerability, we can generate diverse implementations of \texttt{system()} and \texttt{fprintf()} using the same framework for variant generation and integrate a different implementation for each variant of the original code segment. 

\section{Results}
\label{sec:result}
To quantify the effectiveness against Trojans, we perform two major analysis. The first analysis in Subsection \ref{sec:TAR} tries to demonstrate that among the extensive design space of possible trigger conditions within a processor, only a very few can activate in multiple variants. The second analysis in Subsection \ref{sec:det_res} shows the detection and resilience against hardware Trojans that activate in single and multiple variants. Please note that in both analyses, we only considered Trojans that activate in at least one of the variants of the simulated programs to observe the behaviour of the variants under an attack. However, being able to activate a Trojan is not a requirement for our technique unlike pre-deployment verification \cite{chakraborty2009mero}.   

\subsection{Trigger Avoidance Rate Analysis}
\label{sec:TAR}

\subsubsection{Objective and method}
It is critical to understand if the compiler-generated variant-pairs contain trigger conditions that activate in multiple variants. 
To analyze this we execute one of the variants of a program and observe signals with low but non-zero static probability (SP). 
A combination of these signals can be used as trigger inputs for creating Trojans that activates when all the trigger inputs to the Trojan become logic high simultaneously. First, we formulate large number of such Trojans of different input sizes (from 4 to 8) that successfully triggers in one of the variants and observe if those hypothetical Trojans activate again during the execution of other variants of the same program.
To better understand the trend, we formulate the Trigger Avoidance Rate (TAR) for different variant pairs that reflects the percentage of trigger conditions that activates in one variant but stays dormant in the other variant. If $T_{n}$ different valid trigger conditions are found during the execution of  variant $n$ and among them $T_{m}$ trigger conditions are also satisfied in  variant $m$ (including the compare operation), the $TAR$ for that variant-pair $n$-$m$ is found as: $$TAR= \dfrac{(T_{n} -T_{m})}{T_{n}}\times100$$ 
By analyzing $TAR$, we seek to answer the following questions:
a) What is the probability that an arbitrary valid trigger condition will re-trigger in multiple variants?
b) How TAR changes with trigger difficulty of Trojans?
c) How the quality of a variant-pair impacts TAR?

\subsubsection{Simulation process}
We generated three variants from a linear feedback shift register (LFSR) program written in C using the LLVM-based framework. 
We executed these variants in the Plasma processor from OpenCores which is a 32-bit RISC microprocessor \cite{plasma}. We simulated the IP core in Xilinx Vivado and captured the signal states of the internal wires.

\begin{figure}[t!]
\centerline{\includegraphics[width=\columnwidth]{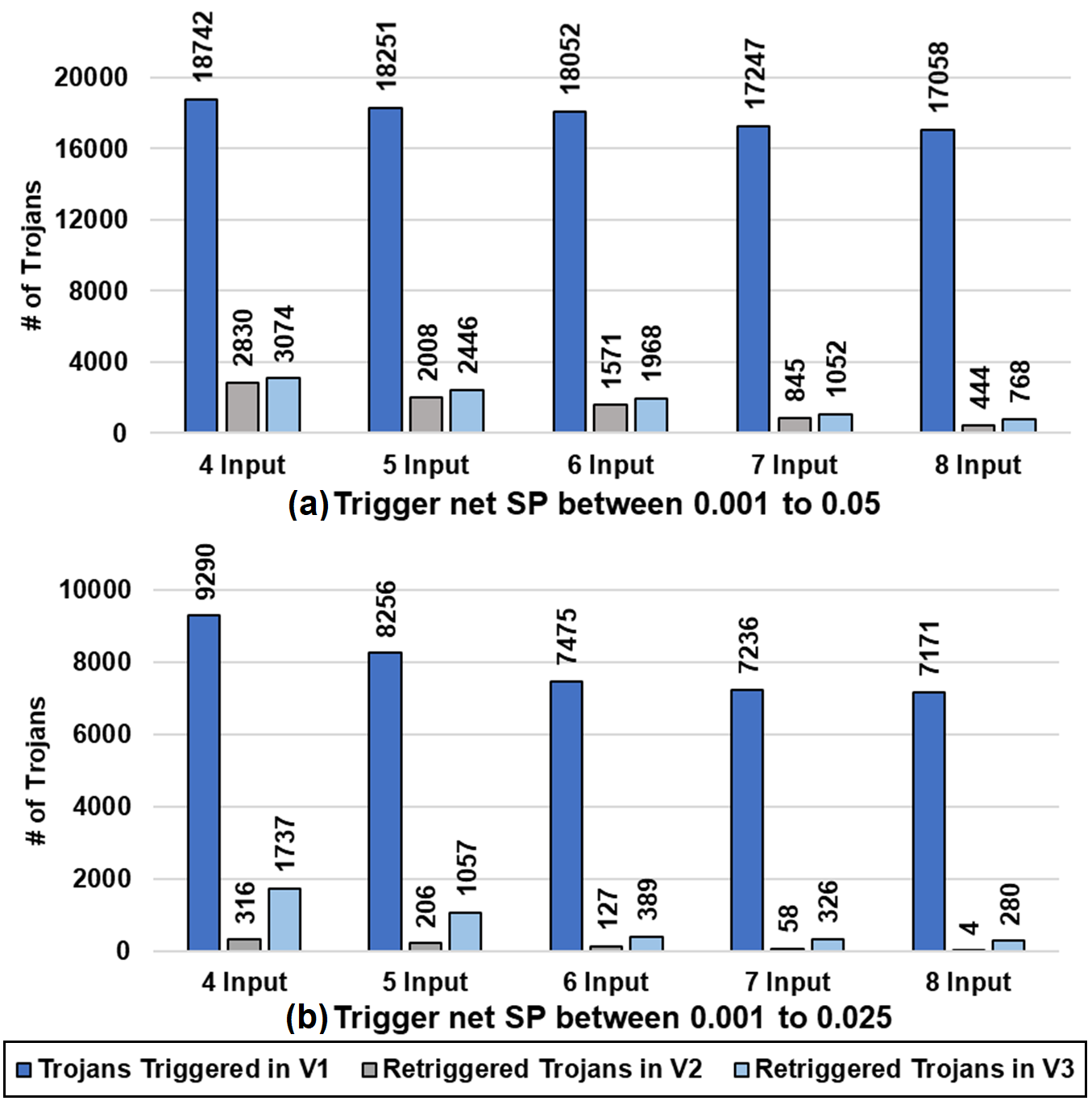}}
\caption{Trojan activation analysis for large number of combinational trigger conditions of various input sizes from 4 to 8. For (a), the considered trigger nets have a maximum static probability is 0.05 and for (b), its 0.025. In both cases, a small fraction of the activated trigger conditions in variant 1 reactivates in variant 2 and variant 3. This number decreases as we consider larger trigger inputs (i.e., complex trigger). For (b), the number of activated triggers in all variants are lower compared to (a) due to the smaller static probability range. }
\label{fig:lfsr1}
\end{figure}

\subsubsection{Analysis of results}
To construct the emulated Trojans we analyzed all the internal signal states from simulation of the Plasma processor during variant-1 execution and calculated the SP of all signals.  
The value of SP ranges from 0.00 to 1.00, with lower values (e.g., $<$0.5) indicating that the signal stays at logic-1 state for shorter amount of time throughout the simulation and vice-versa. 
We divide the low SP signals into two groups, one with SP range between 0.05 to 0.001 and other between 0.025 to 0.001. These nets rarely become logic-1 during variant-1 execution and we can construct Trojan triggers of different input sizes using these nets. 
As shown in Fig. \ref{fig:lfsr1} (a) and (b), for both the SP ranges, we were able to find a large number of combinational trigger conditions of 4, 5, 6, 7, and 8 inputs that activate at least once during variant-1 execution. However, for lower SP range among these two (i.e., Fig. \ref{fig:lfsr1} (b)), the number of rare trigger conditions reduces due to the presence of fewer nets in this SP range that activate in variant-1 (e.g., from 18742 down to 9290 for 4-input triggers). The Trojan space also reduces as we consider larger input size of the triggers from 4 to 8 (i.e., complex trigger condition).

As shown in Fig. \ref{fig:lfsr1} (a), when variant-1 executes, 18742 trigger conditions can be constructed that have 4 trigger nets with  maximum SP of 0.05. However, when we observed those activated Trojans during the execution of variant-2, only 2830 of them activated again. For variant 3, the number is also considereably lower. The number of reactivated Trojans in variant-2 and variant-3 reduces as we consider larger input sizes from 4 to 8. This is also the case for Fig. \ref{fig:lfsr1} (b), where the trigger nets are from signals with SP below 0.025. In this case, the number of reactivated Trojans in variant-2 and 3 are even smaller. 

\begin{figure}[t!]
\centerline{\includegraphics[width=\columnwidth]{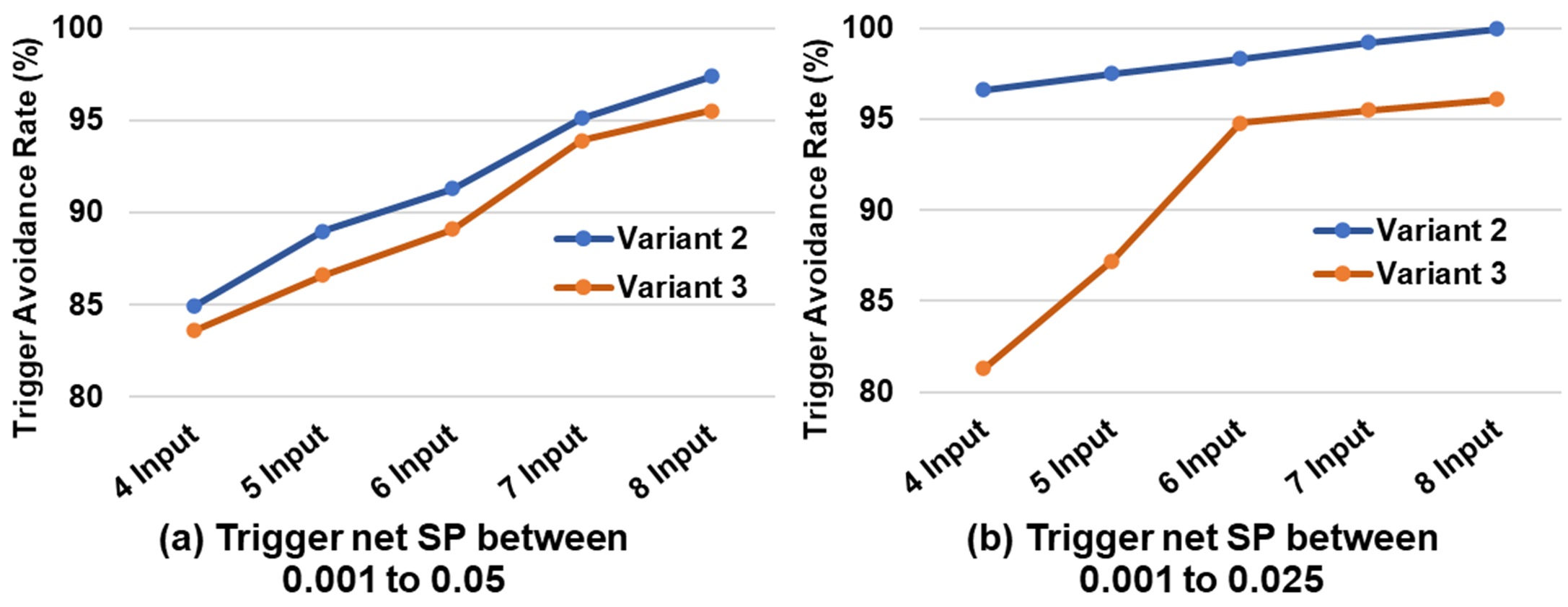}}
\caption{Figure above shows the TAR analysis for the data in Fig. \ref{fig:lfsr1}. 
For both static probability range in (a) and (b), the TAR increases for larger trigger inputs (i.e., complex triggers). Moreover, when considered a lower static probability range for the same trigger input size in (b), the TAR improved for most cases compared to (a). }
\label{fig:lfsr}
\end{figure}

To summarize the findings, we utilize the TAR metric. TAR shows the percentage of Trojans that do not reactivate. Hence, a higher TAR value means the same Trojans are less likely to activate across multiple variants. 
As shown in Fig. \ref{fig:lfsr} (a), the TAR improves for a higher trigger input size. Hence, complex trigger conditions are less likely to reactivate across multiple variants. In Fig. \ref{fig:lfsr} (b), the TAR for signals with lower SP is shown. Compared to (a), the TAR is generally higher for Trojans of the same input size as the signals being considered are less likely to reactivate due to lower SP.  
 Therefore, the LLVM-generated variants are more effective for hard-to-activate Trojans that generally contain larger input size and/or lower SP.

 We also note that variant-2  consistently provides higher TAR compared to variant-3. This observation indicates the benefit of choosing variant pairs that are as diverse as possible. The $VS$ between variant pair 1 and 2 is lower (i.e., more dissimilar) compared to variant pair 1 and 3. Hence, lower $VS$ (lower similarity among variant pairs) is beneficial for avoiding multiple triggering of the same Trojan.

\begin{table*}[t!]
\centering
\caption{Design of Trojans Inserted in the OpenRISC \textit{mor1kx} CPU and their Impact on Variants.  }
\label{tab:exp}
\renewcommand{\arraystretch}{.85}
 \scalebox{0.88}{
\begin{tabular}{|c|c|l|c|l|c|l|c|c|}
\hline
\textbf{\begin{tabular}[c]{@{}c@{}}Trojan \\ Name\end{tabular}} & \textbf{\begin{tabular}[c]{@{}c@{}}Insertion\\ location\end{tabular}} & \multicolumn{1}{c|}{\textbf{Trigger Signal}}                                                                                                                                                                & \textbf{Trigger Type} & \multicolumn{1}{c|}{\textbf{Payload}}                                             & \textbf{\begin{tabular}[c]{@{}c@{}}Activation in \\ V1/V2/V3\end{tabular}} & \multicolumn{1}{c|}{\textbf{\begin{tabular}[c]{@{}c@{}}Output of\\ Variant (Hex)\end{tabular}}} & \textbf{Detection} & \textbf{Tolerence} \\ \hline
\textbf{T1}                                                     & Execution                                                             & \begin{tabular}[c]{@{}l@{}}shift\_msw{[}31:0{]}, exec\_op\_1clk\_i,\\ u\_1clk\_flag\_clear,u\_1clk\_overflow\_set, \\ taking\_1clk\_op\_o, exec\_op\_movhi\_i, \\ exec\_opc\_extsz\_i{[}3:0{]}\end{tabular} & Combinational         & \begin{tabular}[c]{@{}l@{}}Invert multiplexer's \\ selection bit\end{tabular}     & 1/0/0                                                                      & \begin{tabular}[c]{@{}l@{}}V1: \color{red}{3E4564C9}\\ V2: 8FFFFF87\\ V3: 8FFFFF87\end{tabular}              & Yes                & Yes                \\ \hline
\textbf{T2}                                                     & Decode                                                                & opc\_insn                                                                                                                                                                                                   & Sequential            & \begin{tabular}[c]{@{}l@{}}Change immediate \\ field value\end{tabular}           & 1/0/0                                                                      & \begin{tabular}[c]{@{}l@{}}V1: \color{red}{8FFFFF8B}\\ V2: 8FFFFF87\\ V3: 8FFFFF87\end{tabular}              & Yes                & Yes                \\ \hline
\textbf{T3}                                                     & Decode                                                                & fetch\_rfa2\_adr\_i                                                                                                                                                                                         & Sequential            & \begin{tabular}[c]{@{}l@{}}Change ALU logic \\ operation\end{tabular}             & 1/0/0                                                                      & \begin{tabular}[c]{@{}l@{}}V1: \color{red}{DFFFFFDC}\\ V2: 8FFFFF87\\ V3: 8FFFFF87\end{tabular}              & Yes                & Yes                \\ \hline
\textbf{T4}                                                     & Decode                                                                & opc\_alu                                                                                                                                                                                                    & Sequential            & \begin{tabular}[c]{@{}l@{}}Invert immediate \\ selection control bit\end{tabular} & 1/0/0                                                                      & \begin{tabular}[c]{@{}l@{}}V1: \color{red}{8FFEFFDD}\\ V2: 8FFFFF87\\ V3: 8FFFFF87\end{tabular}              & Yes                & Yes                \\ \hline
\textbf{T5}                                                     & Decode                                                                & \begin{tabular}[c]{@{}l@{}}fetch\_rfa1\_adr\_i, fetch\_rfa2\_adr\_i, \\ fetch\_rfb1\_adr\_i, fetch\_rfb2\_adr\_i, \\ fetch\_rfd1\_adr\_i, fetch\_rfd2\_adr\_i\end{tabular}                                  & Combinational         & \begin{tabular}[c]{@{}l@{}}Change immediate\\ field value\end{tabular}            & 1/1/0                                                                      & \begin{tabular}[c]{@{}l@{}}V1: \color{red}{8FFFFF8B}\\ V2: \color{red}{8FFFFF8B}\\ V3: (didn't run)\end{tabular}          & No                 & No                 \\ \hline
\textbf{T6}                                                     & Decode                                                                & opc\_insn                                                                                                                                                                                                   & Sequential            & \begin{tabular}[c]{@{}l@{}}Change immediate \\ field value\end{tabular}           & 1/1/0                                                                      & \begin{tabular}[c]{@{}l@{}}V1: \color{red}{7F3A554A}\\ V2: \color{red}{313D074F}\\ V3: 8FFFFF87\end{tabular}              & Yes                & No                 \\ \hline
\textbf{T7}                                                     & Decode                                                                & dcod\_opc\_setflag\_o                                                                                                                                                                                       & Sequential            & \begin{tabular}[c]{@{}l@{}}Change ALU logic \\ operation\end{tabular}             & 1/1/0                                                                      & \begin{tabular}[c]{@{}l@{}}V1: \color{red}{DFFFFFDC}\\ V2:\color{red}{AFFFFFA9}\\ V3: 8FFFFF87\end{tabular}               & Yes                & No                 \\ \hline
\textbf{T8}                                                     & Decode                                                                & opc\_alu                                                                                                                                                                                                    & Sequential            & \begin{tabular}[c]{@{}l@{}}Invert immediate \\ selection control bit\end{tabular} & 1/1/0                                                                      & \begin{tabular}[c]{@{}l@{}}V1: 8FFFFF87\\ V2: \color{red}{8FFEFFDD}\\ V3: 8FFFFF87\end{tabular}              & Yes                & Yes                \\ \hline
\textbf{T9}                                                     & Execution                                                             & \begin{tabular}[c]{@{}l@{}}exec\_op\_1clk\_i, u\_1clk\_flag\_clear, \\ u\_1clk\_overflow\_set,  taking\_1clk\_op\_o, \\ exec\_op\_movhi\_i, exec\_opc\_extsz\_i{[}3:0{]}\end{tabular}                       & Combinational         & \begin{tabular}[c]{@{}l@{}}Invert multiplexer's \\ selection bit\end{tabular}     & 1/1/0                                                                      & \begin{tabular}[c]{@{}l@{}}V1: 8FFFFF87\\ V2: \color{red}{E9CB4790}\\ V3: 8FFFFF87\end{tabular}              & Yes                & Yes                \\ \hline
\end{tabular}}
\end{table*}

\subsection{Detection and Resilience Analysis}

\label{sec:det_res}
\subsubsection{Objective and method} The goal of this analysis is to understand the effectiveness our solution in detecting and tolerating the activation of Trojans in a single, as well as in multiple variants. 
While the previous analysis showed that arbitrarily inserted hard-to-activate trigger conditions rarely trigger in multiple variants, it is important to discuss the impact of both single and multiple activation. 
To facilitate our analysis, we have handcrafted and inserted Trojans inside a processor by altering its hardware description language. 

\subsubsection{Simulation process}
We selected the OpenRISC \textit{mor1kx} CPU with \textit{Marocchino} pipeline implementation written in Verilog as our target platform for Trojan insertion. The program to be executed on the processor is the Tiny Encryption Algorithm (TEA) written in C. We combined our variant-generation framework with the  LLVM backend for openRISC and created the variant-integrated assembly from the C implementation of TEA. Finally, we used a cross-compiler GCC toolchain “or1k-elf-gcc” to create the executable.
We used Icarus Verilog along with FuseSoC Linux command-line tool to simulate the  executable in the processor.

Since our analysis requires activation of the Trojans during simulation, the trigger circuits are designed based on the internal activity of the processor when the target executable of the TEA program is running. Therefore, first we monitored the waveform of various internal signals of the processor that can serve as valid trigger conditions for the variant-integrated TEA program. Table \ref{tab:exp} shows the location, trigger signals, and trigger type for each Trojan we inserted. For example, the first Trojan T1 is inserted in the execution module of the processor and the second one is inserted in the decoder. A combinational trigger means that the Trojan is activated as soon as the desired trigger value appears on the trigger signals. The sequential triggers are activated only when a desired ``sequence'' of values appear on the trigger signals. Our threat model primarily focuses on Trojans that try to alter the program output, instead of causing a denial of service through arbitrary or persistent corruption. Therefore, the payload signal for each Trojan are designed to cause a transient corruption of different signals that leads to a faulty encryption result generated by the TEA program. 

\subsubsection{Analysis of results} 
The first four Trojans (T1-T4) activate only once and impact the first variant while the other five Trojans (T5-T9) activate in multiple variants. Table \ref{tab:exp} shows the activation of the Trojans in different variants and the corresponding encrypted output. If the encryption output does not match for the first two variants, the Trojan gets detected during the comparison and third variant is automatically executed for majority voting. 
The Trojan is assumed to be tolerated only when the majority voting provides the correct encryption result (i.e., 32'h8FFFFF87). 

 Once \textbf{Trojan T1} activates, the addition operation in the execution module is performed with a corrupted value. As a result, the processor produces an incorrect encrypted output in the first variant (shown in red fonts). On the other hand, the processor generates the correct output for the second variant as the Trojan stays untriggered. After comparing the result of the first two variants, the Trojan was detected. Due to the mismatch in the result, the third variant got executed and produced correct output as T1 did not reactivated.
 Activation of \textbf{Trojan T2} corrupts the immediate value of the instruction that could either be the program's input data or the memory address. During the activation of T2, the immediate field contained input plaintext that led to a corrupted encryption output in the first variant but correct result in the second. The Trojan was detected and tolerated in the earlier manner as it does not trigger during the second and third variant. 
Triggering of \textbf{Trojan T3} changes the next instruction from addition to multiplication leading to a corrupted encryption result. As the Trojan only activated in the first variant and was dormant in second and third variant, it was detected and the correct encryption result was found from majority voting. \textbf{Trojan T4} changes the immediate selection bit of the decoded instruction. During the execution of the first variant the Trojan activates and causes the encryption result to change. For the second and third variant, the Trojan did not trigger and got detected and tolerated.

\textbf{Trojan T5} activated in both the variants and corrupted the encryption output in the same manner. It corrupted the immediate field of the instruction that contains the input plain-text in both the variants. Since the plain-text is identical for all variants, the final impact on the encrypted output remained the same.   Protection against such Trojan can be improved with more diverse variants and judicious transformation of register usage for critical data across variants. \textbf{Trojan T6 and T7} triggers in both first and second variant but unlike T5, they do not corrupt the output in identical manner for both the variants. Therefore, the Trojan is detected due to the mismatch of the output. However, since both the variants generated faulty outputs, correct execution of the third variant could not provide the correct output through majority voting. Such a scenario could be addressed by including more variants (e.g., five or seven). Finally, \textbf{Trojan T8 and T9} are similar to T6 and T7 as they all activate during first and second variant. However, activation of T8 and T9 during first variant execution do not impact the encryption output. Therefore, by executing the third variant, correct encryption results were found.

\begin{table}[t!]
\centering
\caption{Programs and Protected Variables Used to Analyze the Code Size and Runtime Overhead}
\label{tab:prog}
 \scalebox{0.88}{
\begin{tabular}{|c|c|c|c|}
\hline
\textbf{Program   name} & \textbf{Variable   Name}                                                                   & \textbf{Variable  Type}   & \textbf{O/P Size } \\ \hline
\textbf{bitcount}       & n                                                                                          & long   Int                & 4   KB                         \\ \hline
\textbf{CRC32}          & oldcrc32,  charcnt, crc                                                                    & long   Int x 3             & 491   MB                       \\ \hline
\textbf{DES}            & result                                                                                     & long  long hex            & 32   KB                        \\ \hline
\textbf{dijkstra}       & chNode,   qNext                                                                            & long   Int, Int           & 596   KB                       \\ \hline
\textbf{FFT}            & RealOut,   ImagOut                                                                         & float, hex                & 948   KB                       \\ \hline
\textbf{patricia}       & time,addr.s\_addr                                                                          & float,  hex               & 1.3   MB                       \\ \hline
\textbf{qsort}          & \begin{tabular}[c]{@{}c@{}}array{[}i{]}.x,  array{[}i{]}.y, \\ array{[}i{]}.z\end{tabular} & Int                       & 1.5   MB                       \\ \hline
\textbf{rawcaudio}      & outp                                                                                       & unsigned  char            & 102   MB                       \\ \hline
\textbf{rawdaudio}      & outp                                                                                       & Int                       & 74   MB                        \\ \hline
\textbf{sha}            & data, count, digest                                                                        & long   Int, Int, long hex & 2   MB                         \\ \hline
\textbf{susan}          & out                                                                                        & unsigned   char           & 456   KB                       \\ \hline
\textbf{Triple   DES}   & value,   plain                                                                             & char,   long int          & 568   KB                       \\ \hline
\end{tabular}}
\end{table}

\subsection{Overhead Analysis}
\begin{figure}[t!]
\centerline{\includegraphics[width=\columnwidth]{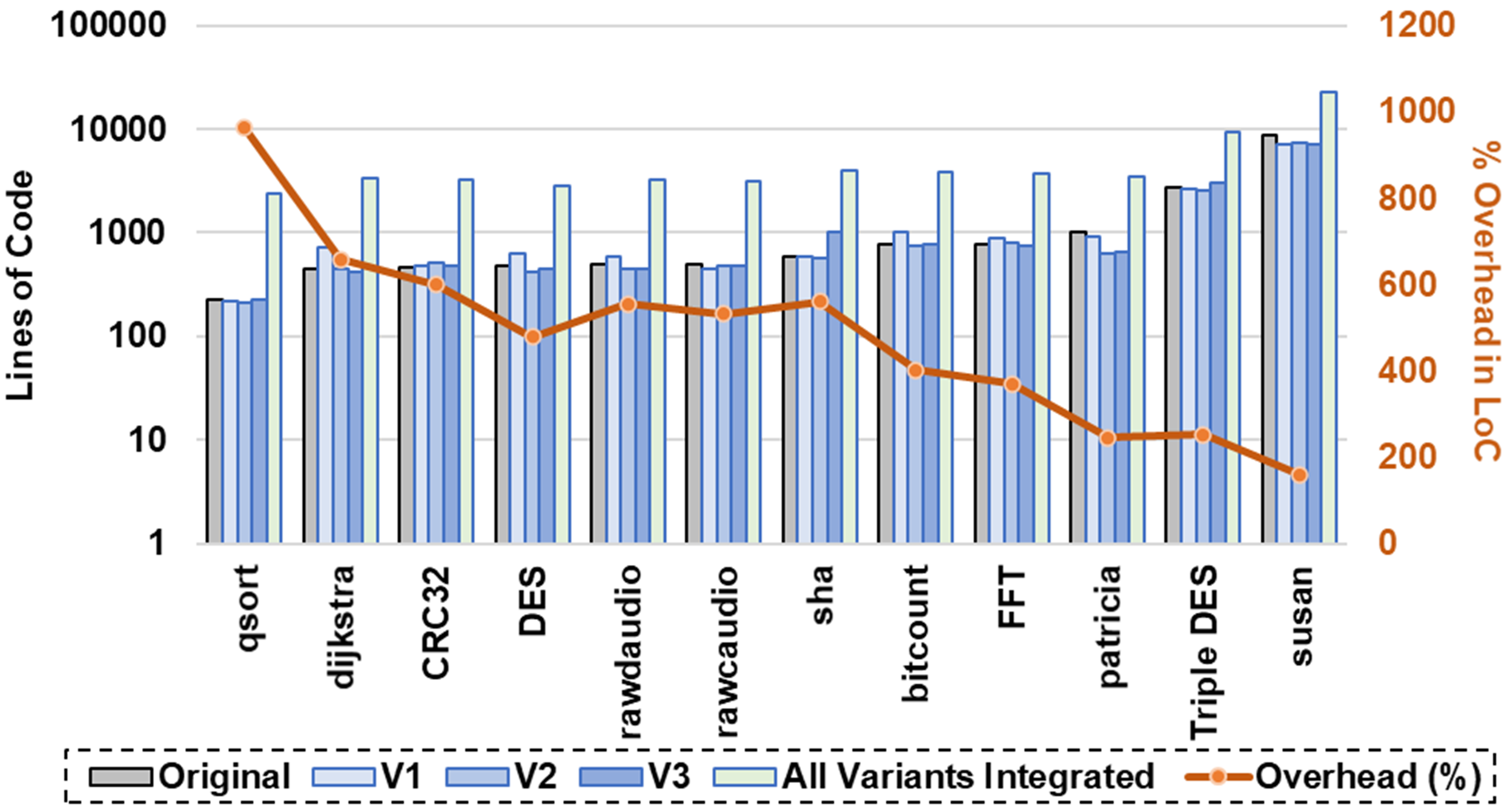}}
\caption{Code size overhead of our variant-integrated approach for several programs. The left vertical axis denotes lines of code (LoC) and the right one denotes \% of increase in LoC for our solution.     }
\label{fig:code_overhead}
\end{figure}

\begin{figure}[t!]
\centerline{\includegraphics[width=\columnwidth]{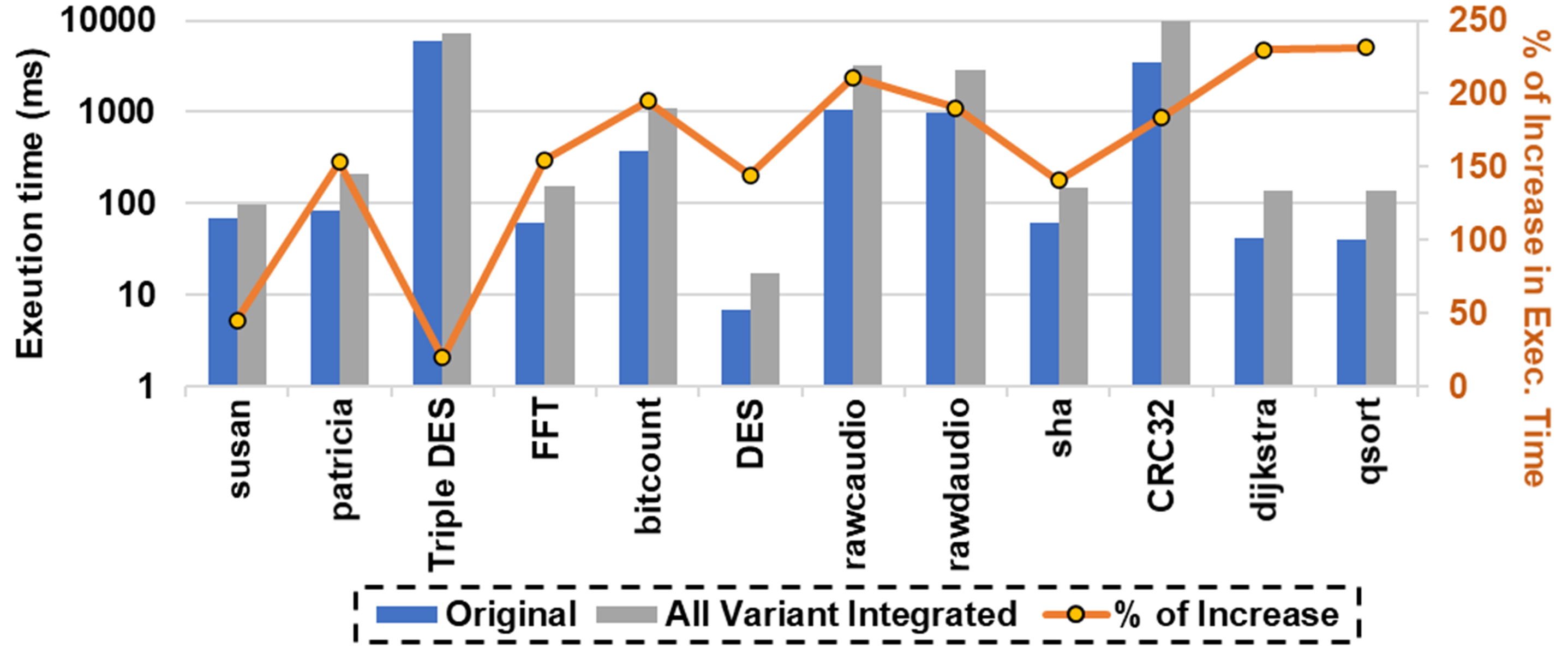}}
\caption{Execution time overhead of our variant-integrated approach for several programs. The left vertical axis denotes execution time and the right one denotes \% of increase in execution time for our solution.   }
\label{fig:run_overhead}
\end{figure}

Fig.~\ref{fig:code_overhead} and Fig.~\ref{fig:run_overhead} present the overhead in code size and execution time due to the variant-based execution of several C programs from the MiBench suite and two implementations of Data Encryption Standard (DES) algorithm. Table \ref{tab:prog}, includes variable names and their types that are protected (i.e., compared after variant execution) in each program. It also includes the size of the file where these variables are stored as program output for comparison. 
The code overhead is calculated by comparing the lines of code in the assembly version of the original program with the variant-integrated one. The variant-integrated program includes codes for three variants and system calls to compare the variables.  As shown in Fig. \ref{fig:code_overhead}, using the left vertical axis, we presented the code size of the original program, individual code size of the three variants, and code size of the variant-integrated solution that contains three variant and additional codes to execute them and compare results. Since the programs are very different with respect to their code size, we used a logarithmic scale for the Lines of Code (LoC). We can observe that the original program and its three diverse variants have similar LoCs. Therefore, our LLVM-based framework is generating variants with similar code sizes, yet diverse with respect to opcode and operands. 
Using the right vertical axis, we presented the percentage of increase in LoC from the original program to the variant-integrated one. Please note that the overhead reduces gradually as we consider programs with large code size. Assuming the three variants are of similar size, our framework should increase the code size by 200\% for any program, but the additional code for integration of variants and comparing the result leads to different overhead depending on the size of the original program. For a small program like qsort with 227 LoC, the additional code for integration (i.e., 1751 LoC) impacts the overhead significantly. However, for a larger program like susan	with 8784 LoC the overhead is much lower.

To calculate the execution time, we used performance analyzing tool in Linux called perf. We used the 12-core Ryzen 3900x CPU. Fair comparison of the execution time overhead requires us to use a single core for executing the original program as well as the variant-integrated one. Hence, we disabled all the cores except one to ensure that the program would run only in a single core. The perf tool was used to run each program 10 times and obtain the average execution time. As shown in Fig. \ref{fig:run_overhead}, using the left vertical axis, we presented the execution time of the original program and the variant-integrated counterpart that executes two variant and compares their results. The overhead could vary depending on  the difference between the runtime of the original program and the runtime of two variants. This could be observed by relating the code overhead with runtime overhead. Programs with low code overhead such as susan and Triple DES incurred lower runtime overhead as well. However, the runtime overhead could also depend on which variables within the program are compared. As shown in Table  \ref{tab:prog}, three different variables are protected in CRC32 that leads to two large files being compared after two variants executed, requiring additional execution time. For applications where such overhead is not admissible, we can only protect a sensitive segment of the program, since an attacker is more likely to include a Trojan to subvert critical operations.

\section{Discussion}
\label{sec:disc}

\subsection{Trojan Insertion based on the Compare Function}
We note that if an attacker is aware of our proposed defense, they may craft a Trojan that will disable the compare logic. However, an attacker would have to enumerate the possible compare instruction instances that can be compiled and executed on the processor to ensure our defense is bypassed.  Several variants can be crafted of the compare function -- CMP, SUB, XOR, etc. If the attacker chooses to corrupt all possible implementations of compare, the number of impacted instructions would increase the detectability of the Trojan as the corruption across original code variants would be high. As a result, we can also run the compare portion of the code through the variant generation and dynamically choose among a valid set of compare functions during runtime.

\subsection{Other Variants Integration Techniques}

In our methodology section, we proposed to use \texttt{system()} function call to run our executable variants \cite{syscall}. The variant execution using \texttt{system()} in the main program is a sequential operation. It implies that the execution order after \texttt{system()} call will stall until the variant execution completes and returns to the main program. Alternatively, we can also use the \texttt{execl()} function call From GNU C \texttt{unistd.h} library \cite{execl}. 
Using \texttt{execl()}, we can run our variants as multiple child processes.
This will enable us to run multiple child processes in parallel in a multi-core system. We can also run each variant in specific cores by assigning the CPU affinity for each variant process. Such parallel execution of variants can help reduce the runtime overhead significantly.



\section{Conclusion}
\label{sec:conc}
Existing countermeasures against hardware Trojans are generally inapplicable for COTS components.
In this paper, we have drawn attention to this serious problem by discussing the limitations of the existing solutions when it comes to COTS ICs. We have presented a novel solution and associated toolflow for runtime Trojan detection and resilience in untrusted COTS processors through the creation and execution of Trojan-aware software variants. The proposed work relies purely on judicious modification of software and, unlike existing solutions, does not require hardware support or architectural changes. We have shown that by selecting diverse variants using the $VS$ metric, the probability of simultaneous activation of a Trojan in multiple variants can be drastically reduced. We demonstrated the detection and resilience capability of our approach using several sample Trojans implemented in a RISC processor. Even though the proposed solution incurs delay in execution time due to the need to perform multiple version of the program and comparing the results, it is one of the very few solutions that could be used for untrusted COTS processors. 

The proposed paradigm can be extended in three major ways. The solution could be advanced to detect and tolerate more diverse Trojan payloads such as persistent corruption and information leakage. Currently our framework primarily addresses payloads with transient corruption. Second, static diversity analysis for a variant pair might differ during execution for COTS with out-of-order execution, speculative execution or similar optimizations. We can extend the similarity metric to account for such behavior in COTS IC.     
Finally, the code size and execution time overhead is also significant.  
Hence, future research should involve identification of the compiler pass sequence that generates diverse variants with lower overhead.     

\ifCLASSOPTIONcaptionsoff
  \newpage
\fi

\bibliographystyle{IEEEtran-style}
\bibliography{IEEEexample}

  \par
\end{document}